\documentclass{article}
\usepackage[final]{graphics}
\usepackage{amsfonts,amsbsy}

\def\empile#1\over#2{\mathrel{\mathop{\kern 0pt#1}\limits_{#2}}}

\newcommand{\sll}{\raise.15ex\hbox{$/$}\kern-.43em\hbox{$l$}}
\newcommand{\slepsilon}{\raise.15ex\hbox{$/$}\kern-.53em\hbox{$\epsilon$}}
\newcommand{\slvarepsilon}{\raise.15ex\hbox{$/$}\kern-.53em\hbox{$\varepsilon$}}

\newcommand{\slL}{\raise.15ex\hbox{$/$}\kern-.53em\hbox{$L$}}
\newcommand{\slP}{\raise.15ex\hbox{$/$}\kern-.53em\hbox{$P$}}
\newcommand{\slp}{\raise.1ex\hbox{$/$}\kern-.63em\hbox{$p$}}
\newcommand{\slq}{\raise.1ex\hbox{$/$}\kern-.63em\hbox{$q$}}
\newcommand{\slv}{\raise.1ex\hbox{$/$}\kern-.63em\hbox{$v$}}
\newcommand{\slR}{\raise.15ex\hbox{$/$}\kern-.53em\hbox{$R$}}
\newcommand{\slQ}{\raise.15ex\hbox{$/$}\kern-.53em\hbox{$Q$}}
\newcommand{\slK}{\raise.15ex\hbox{$/$}\kern-.53em\hbox{$K$}}
\newcommand{\slk}{\raise.15ex\hbox{$/$}\kern-.53em\hbox{$k$}}
\newcommand{\slSigma}{\raise.15ex\hbox{$/$}\kern-.53em\hbox{$\Sigma$}}
\newcommand{\slcalP}{\raise.15ex\hbox{$/$}\kern-.63em\hbox{$\cal P$}}
\newcommand{\slA}{\raise.15ex\hbox{$/$}\kern-.73em\hbox{$A$}}
\newcommand{\slbfA}{\raise.15ex\hbox{$/$}\kern-.73em\hbox{${\imb A}$}}
\newcommand{\slpartial}{\raise.15ex\hbox{$/$}\kern-.53em\hbox{$\partial$}}

\font\tenimbf=cmmib10 at 10pt
\font\sevenimbf=cmmib10 at 7pt
\font\fiveimbf=cmmib10 at 5pt
\newfam\imbf
\textfont\imbf=\tenimbf
\scriptfont\imbf=\sevenimbf
\scriptscriptfont\imbf=\fiveimbf
\def\imb{\fam\imbf\tenimbf}

\begin{document}

\title {\bf Dilepton production from\\ 
the Color Glass Condensate}

\author{Fran\c cois Gelis$^{(1)}$ and Jamal Jalilian-Marian$^{(2)}$}
\maketitle
\begin{center}
\begin{enumerate}
\item Laboratoire de Physique Th\'eorique\\
B\^at. 210, Universit\'e Paris XI\\
91405 Orsay Cedex, France
\item Physics Department\\
         Brookhaven National Laboratory\\
         Upton, NY 11973, USA
\end{enumerate}
\end{center}

 \begin{abstract} 
We consider dilepton production in high energy proton-nucleus (and very
forward nucleus-nucleus) collisions.
Treating the target nucleus as a Color Glass Condensate and describing the
projectile proton (nucleus) as a collection of quarks and gluons as in 
the parton model, we calculate the differential cross section for dilepton 
production in quark-nucleus scattering and show that it is very sensitive 
to the saturation scale characterizing the target nucleus. 
 \end{abstract}
\vskip 4mm

\section{Introduction}

Perturbative QCD predicts a sharp rise in the number of gluons
per unit area and rapidity in a proton/nucleus at high energy 
(small $x$) \cite{bfkl}. This however would lead to violation of unitarity of 
hadronic cross sections at high energies. High gluon density and 
gluon recombination effects \cite{glrmq} 
are believed to be responsible for taming this growth and restoration 
of unitarity. It has been suggested that gluons at small $x$ can be
described by a strong classical field \cite{mv} and that weak coupling, 
semi-classical methods can be applied to describe the physics of 
dense gluonic systems, such as a proton or nucleus at high energies.
This dense system of gluons (the Color Glass Condensate) is characterized 
by a saturation momentum $Q_s (x)$ which grows fast with energy and 
rapidity.

There has been much work done in order to investigate the properties
of the Color Glass Condensate in DIS electron-proton and electron-nucleus,
as well as proton-nucleus and nucleus-nucleus collisions \cite{sat}. 
While the saturation effects in protons at current energies are far from 
being established \cite{gbw}, the situation in high energy nucleus-nucleus 
collisions
is more intriguing. The saturation model seems to work reasonably well
at RHIC \cite{hisat} even though there are a lot of open questions which 
need to be addressed before one can claim that the saturation model describes
high energy heavy ion collisions at RHIC quantitatively. The role of 
final state interactions, thermalization, etc. is still to be understood 
\cite{finsat}.

In a recent set of publications \cite{pa,gjm}, we proposed that high energy
proton-nucleus collisions at forward rapidities at RHIC may be an 
ideal place in order to investigate the Color Glass Condensate and 
the saturation model. By considering proton-nucleus collisions in the
forward rapidity region, one can avoid most, if not all, of the 
complications present in a nucleus-nucleus collision. 
Previously, we calculated (real) photon production rate in $p-A$ 
collisions \cite{gjm} and showed that it is very sensitive to saturation 
effects in the nucleus. 

In this work, we consider dilepton (virtual photon) production 
in $p-A$ and show that dileptons provide a more versatile probe of
the saturation model than photons. Furthermore, photons are
notoriously difficult to measure in a collider environment. One needs
to define isolation criteria in order to separate photons from
different sources which greatly reduces the production rates in 
addition to introducing theoretical ambiguities in defining
the isolation criteria. The use of factorization theorems, well 
established for high $p_\perp$ {\it inclusive} photon production, may
be questionable for isolated photons. By considering dilepton
production, one can avoid most of these experimental and theoretical 
difficulties \cite{bgkp}.

We briefly review our formalism and the differences between real and
virtual photon cross sections in Section $2$. In section $3$,
we consider the diffractive cross section and show that it vanishes.
We consider the inclusive cross section in Section $4$ and 
derive the differential cross section for dilepton production 
in $p-A$ collisions $d\sigma /dz\,dM^2\,d^2{\imb k_\perp}$, where $M^2$ and ${\imb k_\perp}$
are the dilepton invariant mass and transverse momentum while $z$ is
its fractional energy. We end by discussing our results and the
experimental signatures of saturation effects.

\section{Real vs. virtual photon production}
In order to reuse some parts of the calculation we already performed for
real photon production \cite{gjm}, we start by a section highlighting the main
differences between real photon and lepton pair production, as well as the
common aspects.

\begin{figure}[ht]
\centerline{\resizebox*{!}{2.5cm}{\includegraphics{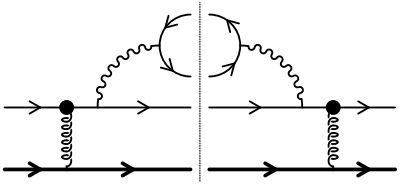}}}
\caption{\label{process} A typical contribution to the cross-section for
lepton pair production in pA collisions. The black dot denotes the
re-summed interactions of the incoming quark with the classical color
field of the nucleus. This is the square of the term where the photon is
emitted after the scattering on the nucleus. There is also a term where
the photon is emitted first, and interferences thereof, not represented
here.}
\end{figure}

We want to calculate the amplitude for the elementary process
\begin{equation}
q({\imb p})+A \to q({\imb q})+l^+({\imb k_1})+l^-({\imb k_2})+X
\end{equation}
where a quark entering in the color field of a nucleus emits a lepton
pair $l^+l^-$. In terms of {\sl in-} and {\sl out-} states, this
amplitude reads:
\begin{equation}
\left<q({\imb q})l^+({\imb k_1})l^-({\imb k_2})_{\rm out}|q({\imb
p})_{\rm in}\right>
=\big<0_{\rm out}\big|b_{\rm out}({\imb
q})b^{\dagger}_{\rm in}({\imb p})
c_{\rm out}({\imb k_2})d_{\rm out}({\imb k_1})\big|0_{\rm in}\big>\; ,
\end{equation}
where $b^\dagger$ is the creation operator for a quark, while
$c^\dagger$ and $d^\dagger$ respectively create a lepton and an
anti-lepton. Applying the LSZ reduction formula to this amplitude \cite{iz}, 
one obtains the following expression in terms of the fermionic fields:
\begin{eqnarray}
&&\big<0_{\rm out}\big|b_{\rm out}({\imb
q})b^{\dagger}_{\rm in}({\imb p})
c_{\rm out}({\imb k_2})d_{\rm out}({\imb k_1})\big|0_{\rm in}\big>=
\int d^4x\, d^4y\, d^4z_1\, d^4z_2\nonumber\\
&&\qquad\times e^{i(q\cdot x-p\cdot y)}\,
e^{i(k_1\cdot z_1 + k_2\cdot z_2)}
\overline{u}({\imb q})\overline{w}({\imb k_2})
(i \stackrel{\rightarrow}{\slpartial}_x -m)
(i \stackrel{\rightarrow}{\slpartial}_{z_2} -m)\nonumber\\
&&\qquad\times
\big<0_{\rm out}\big|{\rm T}\psi(x)\overline{\psi}(y)
\overline{\Psi}(z_1)\Psi(z_2)
\big|0_{\rm in}\big>
(i\stackrel{\leftarrow}{\slpartial}_y +m)
(i\stackrel{\leftarrow}{\slpartial}_{z_1} +m)
v({\imb k_1}) u({\imb p})\; ,\nonumber\\
&&
\end{eqnarray}
where $\psi$ is the quark field, $\Psi$ the leptonic field, $u$ a quark
free spinor, $w$ a lepton free spinor and $v$ an anti-lepton free
spinor. Note that we have approximated all the renormalization constants
by $1$ since we are going to compute only the lowest order in the
couplings $\alpha_{\rm em}$ and $\alpha_{_{S}}$. Therefor this amplitude
is made of a quark line and a leptonic line, connected by photons. The
quark line interacts with the background color which is used to describe
the high energy nucleus. At lowest order in the electro-magnetic and
strong coupling constants, only one bare photon connects the two
fermionic lines, as illustrated in figure (\ref{process}).

The part of the diagram that describes the scattering of the quark in
the color field of the nucleus is identical to what we have already
calculated for the case of photon production. Namely, the photon can be
attached before or after the scatterings, but the terms where the photon
is attached between scatterings of the quark on the nucleus are
suppressed by inverse powers of the center of mass energy
$\sqrt{s}$. Therefore, compared to the photon production case, we need
only to replace the photon polarization vectors of the produced photon
by the propagator of the (now virtual) photon, its coupling to the
leptonic line and the spinors of the $l^+l^-$ pair. This amounts to the
following substitution\footnote{We have indicated the $i\epsilon$
prescription for the photon propagator, but it is in fact irrelevant
here since the photon must have an invariant mass squared $k^2$ larger
than $4m_l^2$ because of the threshold for the production of a pair of
leptons with mass $m_l$.}
\begin{equation}
\epsilon_\mu({\imb k})\epsilon_\nu^*({\imb k})\to
{{g_{\mu\rho}}\over{k^2+i\epsilon}}
{{g_{\nu\sigma}}\over{k^2-i\epsilon}}
L^{\rho\sigma}(k_1,k_2)\; ,
\end{equation}
where $k\equiv k_1+k_2$ is the 4-momentum of the virtual photon, and
where $L^{\rho\sigma}(k)$ is the discontinuity of the one-loop leptonic
contribution to the photon polarization tensor (i.e. the loop on the
upper part of the diagram of figure \ref{process} - only its
discontinuity is needed since the leptons are produced on-shell).  

An important simplification we used in the photon case was that the sum
over the photon polarizations turned the product $\epsilon_\mu({\imb
k})\epsilon_\nu^*({\imb k})$ into $-g_{\mu\nu}$ (up to terms
proportional to $k_\mu$ that do not contribute thanks to Ward
identities). This property led to a dramatic simplification of the Dirac
algebra involved in the calculation of the photon production
cross-section.  A priori, the object $L^{\rho\sigma}(k_1,k_2)$ is not
proportional to $g^{\rho\sigma}$, which means that the {\sl fully
differential} lepton pair production cross-section has a more
complicated Lorenz structure. If however we assume that one
reconstructs the virtual photon 4-momentum from the momenta $k_1,k_2$ of the
components of the lepton pair, the same simplification occurs. Indeed,
if one integrates over the lepton momentum inside the leptonic tensor
$L^{\rho\sigma}$, keeping the sum $k=k_1+k_2$ fixed, we obtain the
following result:
\begin{equation}
L^{\rho\sigma}={2\over 3}\alpha_{\rm em}(g^{\rho\sigma}k^2-k^\rho
k^\sigma)\; .
\end{equation}
Therefore, for the cross-section $d\sigma/d^4k$, we have the same
simplification as in the photon production case.

The relation between the differential cross-section and the amplitude
can therefore be written as
\begin{equation}
d\sigma={{d^4k}\over{(2\pi)^4}}{{d^3{\imb q}}\over{(2\pi)^3 2q_0}}
{1\over{2p^-}}{{2\alpha_{\rm em}}\over{3 k^2}} {\cal M}^\mu({\imb
p}|{\imb q}k){\cal M}_\mu^*({\imb p}|{\imb q}k)
2\pi\delta(q^-+k^--p^-)\; , 
\end{equation} 
where ${\cal M}^\mu$ is the amplitude for the production of a virtual
photon, amputated of its external legs, and from which the factor
$2\pi\delta(q^-+k^--p^-)$ has been removed. Explicitly, we have:
\begin{eqnarray}
&&{\cal M}^\mu({\imb p}|{\imb q}k)= -ie_q\,\overline{u}({\imb q})\bigg[ 
{\gamma^- ({\slp} - {\slk} + m) {\gamma^\mu} \over (p-k)^2 -m^2} +
{{\gamma^\mu}\;({\slq} + {\slk} + m)\gamma^- \over (q+k)^2 -m^2}
\bigg]u({\imb p})\nonumber\\
&&\qquad\qquad\times\int d^2{\imb x}_\perp
e^{i({\imb q}_\perp+{\imb k}_\perp-{\imb p}_\perp)\cdot{\imb x}_\perp}
\Big(U({\imb x}_\perp)-1\Big)\; ,
\label{eq:finalamp}
\end{eqnarray}
where $e_q$ is the electrical charge of the quark and where $U({\imb
x}_\perp)$ is a matrix in the fundamental representation of $SU(N_c)$
that represents the interactions of the quark with the classical color
field of the nucleus:
\begin{equation}
U({\imb x}_\perp) \equiv {\rm T} \exp \bigg \{-ig^2 \int^{+\infty}_{-\infty}
d z^- {1 \over {\nabla^2_\perp}} \rho_a (z^-,{\imb z}_\perp) t^a \bigg\}
\label{eq:Udef}
\end{equation}
with $t^a$ in the fundamental representation, and where
$\rho_a(z^-,{\imb z}_\perp)$ is the density of color sources in the
nucleus. The color averages over the distribution of hard color sources
in the nucleus are identical to the case of photon production, and can
be found in section 3 of \cite{gjm}.

The factor ${\cal M}^\mu {\cal M}_\mu^*$ in the cross-section differs
from the factor we denoted $|{\cal M}|^2$ in our calculation of photon
production only in the fact that we must not assume $k^2=0$ in the
calculation. In fact, in this quantity, only the factor $\left<{\rm
tr}(L^\dagger L)\right>_{\rm spin}$ that contains the Dirac algebra is
affected by this change. Its new value is now:
\begin{eqnarray}
\left<{\rm tr}(L^{\dagger} L)\right>_{\rm spin}&=&
16m^2\left[
{{p^-{}^2}\over{D_q^2}}
+{{q^-{}^2}\over{D_p^2}}
-{{k^-{}^2}\over{D_p D_q}}
\right]\nonumber\\
&&+8(p^-{}^2+q^-{}^2)\left[
{{2p\cdot q}\over{D_p D_q}}
-{1\over{D_p}}-{1\over{D_q}}
\right]\nonumber\\
&&+8k^2\left[
{{p^-{}^2}\over{D_q^2}}
+{{q^-{}^2}\over{D_p^2}}
+{{(p^-+q^-)^2}\over{D_p D_q}}
\right]\; ,
\end{eqnarray}
where $m$ is the mass of the quark and where we denote $D_p\equiv
(p-k)^2-m^2=-2p\cdot k +k^2$ and $D_q\equiv (q+k)^2-m^2=2q\cdot
k+k^2$. It is trivial to check that it reduces to the value found in
\cite{gjm} if we set $k^2=0$. Having in mind the fact that the quark
comes from the wave-function of a proton, we neglect the mass of the
quark in the following.

\section{Diffraction}
Like in \cite{gjm}, we can first study the case of diffractive
dilepton production, as the kinematics is simpler. Let us just remember
that the diffractive cross-section is obtained by performing the average
over the distribution of nuclear color sources before squaring the
amplitude. This implies that no net transverse momentum is exchanged
between the nucleus and the quark, i.e. ${\imb p}_\perp={\imb
q}_\perp+{\imb k}_\perp$. If one evaluates the factor $\left<{\rm
tr}(L^{\dagger} L)\right>_{\rm spin}$ with this kinematical constraint,
one obtains a vanishing result if we neglect the mass $m$ of the quark:
\begin{equation}
\left<{\rm
tr}(L^{\dagger} L)\right>_{\rm spin}^{\rm diff}=0\; .
\end{equation}
This is in fact similar to the case of real photon production: the
absence of transverse momentum exchange between the quark and the
nucleus prevents the emission of the virtual photon.

\section{Inclusive cross-section}
If we do not require a diffractive process on the nuclear side, we just
have to perform the average over colors sources after squaring the
amplitude. The details of this procedure are given in section 6 of
\cite{gjm}. We obtain the following expression for the differential
cross-section:
\begin{eqnarray}
&&d\sigma_{\rm incl}={{d^4k}\over{(2\pi)^4}}{{d^3{\imb q}}\over{(2\pi)^3 2q_0}}
{{e_q^2 \pi R^2}\over{2p^-}}{{2\alpha_{\rm em}}\over{3 k^2}}
\left<{\rm
tr}(L^{\dagger} L)\right>_{\rm spin}\nonumber\\
&&\qquad\qquad\qquad\times
2\pi\delta(q^-+k^--p^-)
C({\imb p}_\perp-{\imb q}_\perp-{\imb k}_\perp)\; ,
\end{eqnarray}
where we define again
\begin{equation}
C({\imb l}_\perp)\equiv \int d^2{\imb x}_\perp e^{i{\imb
l}_\perp\cdot{\imb x}_\perp}
\left<U(0)U^\dagger({\imb x}_\perp)\right>_\rho\; .
\label{eq:C-def}
\end{equation}
Like in the case of real photon production, all the information about the
nature of the medium crossed by the quark (in particular, all the
dependence on the saturation scale $Q_s$) is contained in this function
$C$.

At this point, it is useful to introduce the longitudinal momentum
fraction of the virtual photon $z\equiv k^-/p^-$, as well as the total
transverse momentum transfer between the nucleus and the quark ${\imb
l}_\perp\equiv {\imb q}_\perp+{\imb k}_\perp$. The phase space $d^4k$ of
the lepton pair can be rewritten as $d^4k={1\over 2}d(M^2)(dz/z)
d^2{\imb k}_\perp$, while the phase space of the outgoing quark can be
written as $d^3{\imb q}/(2\pi)^3 2q_0={1\over
2}(dq^-/q^-)\theta(q^-)d^2{\imb l}_\perp$ (implicitly, $q^+={\imb
q}_\perp^2/(2q^-)$). In terms of these new variables, the inclusive 
differential cross-section reads (for unit electric charge of quark)
\begin{eqnarray}
&&\!\!{1 \over \pi R^2}
{d\sigma_{\rm incl}^{q\,A\rightarrow q\,l^+l^-\,X}
\over dz\,d^2{\imb k}_\perp\,d\log M^2}=
{{2 \alpha^2_{\rm em}}
\over{3\pi}}
{{d^2{\imb l}_\perp}\over{(2\pi)^4}} C({\imb l}_\perp)\nonumber\\
&&\qquad\times
\left\{
\left[{{1+(1-z)^2}\over{z}}\right]
{{z^2{\imb l}_\perp^2}\over{[{\imb k}_\perp^2+M^2(1-z)]
[({\imb k}_\perp-z{\imb l}_\perp)^2+M^2(1-z)]}}\right.\nonumber\\
&&\qquad\left.
- z(1-z)\,M^2\left[{1\over{[{\imb k}_\perp^2+M^2(1-z)]}}-{1\over{[({\imb
k}_\perp-z{\imb l}_\perp)^2+M^2(1-z)]}}\right]^2
\right\}\; .\nonumber\\
&&
\label{eq:main}
\end{eqnarray}
This is our main result. It gives the differential cross section for
inclusive production of dileptons in high energy quark-nucleus
collisions and includes all the high gluon density effects in the
nucleus. All the information about the high gluon density effects in the
nucleus is contained in the function $C({\imb l}_\perp)$ \cite{gjm,gp}. 
This function behaves as $1/{\imb l}^4_\perp$ in the 
${\imb l}_\perp \gg Q_s$ 
(perturbative) region and like $1/{\imb l}^2_\perp$ at 
${\imb l}_\perp \sim Q_s$. For ${\imb l}_\perp \ll Q_s$, it is almost
flat. Furthermore, the value of ${\imb l}_\perp $ where the slope of the
cross section changes strongly depends on rapidity. This slow down
happens at higher values of ${\imb l}_\perp $ in the forward rapidity 
region.

This expression reduces to the one found in \cite{gjm} for real 
photon production if we take the limit $M^2\to 0$ as it must. The main 
difference compared to the production of a real photon is the fact that 
the collinear singularities at ${\imb k}_\perp=z{\imb l}_\perp$ and 
${\imb k}_\perp=0$ are now screened by the invariant mass squared of 
the lepton pair, via the term $M^2(1-z)>0$.

In order to relate (\ref{eq:main}) to proton-nucleus collisions, we
will need to convolute (\ref{eq:main}) with the quark distribution
function in a proton using collinear factorization theorem. Explicitly,
\begin{eqnarray}
{d\sigma_{\rm incl}^{p\,A\rightarrow q\,l^+l^-\,X}
\over dz\,d^2{\imb k}_\perp\,d\log M^2} \sim \int\, dx \, q(x,Q^2_f) \,
{d\sigma_{\rm incl}^{q\,A\rightarrow q\,l^+l^-\,X}
\over dz\,d^2{\imb k}_\perp\,d\log M^2}
\label{eq:conv}
\end{eqnarray}
Furthermore, one will need to convolute the above cross section
with a quark/ha\-dron or quark/jet fragmentation function if one is
interested in measuring both the outgoing hadron/jet as well as the
dilepton. Otherwise, one can do the ${\imb l}_\perp$ integration above
to get the $pA\rightarrow l^+\,l^-\,X$ differential cross section.

Experimentally, one will be able to study $M^2$, $k_\perp^2$ and rapidity 
($z$) dependence of the dilepton production cross section in $p-A$
collisions at RHIC in the near future. Here, we outline our qualitative
predictions from the Color Glass Condensate picture of a nucleus at
high energy which will be straightforward to verify/falsify at RHIC. 

First, as compared to the standard leading twist perturbative QCD,
we expect that the partonic level cross section $d\sigma/dy\,d^2{\imb l_\perp}$ 
will change its behavior from $1/l_\perp^4$ to $1/l_\perp^2$ for $l_\perp\sim Q_s$ and an even flatter behavior at smaller $l_\perp$, at fixed rapidity.
Convoluting the partonic cross section with parton structure functions
in order to get the proton-nucleus cross section will change the power 
of $l_\perp$. Nevertheless, we expect the difference in the power of $l_\perp$
to be observable even after the convolution \cite{lt}.

Second, the change of the slope of the cross section from $1/l_\perp^4$
to $1/l_\perp^2$ will happen at a higher transverse momentum in the forward
rapidity region than the mid-rapidity region. Indeed, the saturation scale
of the nucleus near the fragmentation region of the proton is much larger
than the value $Q_s^2 \sim 1-2$GeV$^2$ usually quoted at mid-rapidity \cite{kln}.
The growth of the saturation scale with energy is known from
DIS experiments at HERA \cite{gbw} and heavy ion collisions at 
RHIC \cite{kln}.

For the same reason as above, transverse momentum broadening of the jet+dilepton system will
depend on its rapidity: it will be larger at forward rapidities.
This broadening proportional to $Q_s$ adds up to the broadening  due to initial
"intrinsic transverse momentum" of the incoming quark. This effect has been
neglected in our final result by setting ${\imb p}_\perp=0$. All it would take 
to keep both effects simultaneously would be to keep ${\imb p}_\perp\not=0$ in the calculation.

A more quantitative investigation of our results is beyond the
scope of this work and will be pursued elsewhere. Nevertheless, we
would like to point out that inclusion of the standard leading order
pQCD diagrams \cite{esw} such as dilepton production via 
quark-anti-quark annihilation\footnote{In principle, one will have to
include the effects of high gluon densities on sea quarks. However, this
will have another factor of $\alpha_s$ and therefore is higher order} 
and from direct (virtual) photon diagrams will be required for numerical
accuracy. Since these diagrams are not effected by the strong classical field
of the nucleus and do not interfere with the diagrams considered 
in this work, one can just add their contribution to our results. This
way, one will have the full LO [$O(\alpha_{em}^2)$] and 
NLO [$O(\alpha_s\,\alpha_{em}^2$] dilepton production cross section
in $p-A$ including the high gluon density effects in the nucleus \cite{eks}.

We would like to emphasize that our result (\ref{eq:main}) can also
be used for heavy ion collisions in the very forward rapidity region
where valence quarks are the dominant partons in the projectile nucleus.
The only difference with $p-A$ is that one would then need to convolute
our cross section with the quark distribution function in a nucleus 
rather than a proton. This may make it possible to extract the shadowing 
function for quarks (at large $x$) in the projectile nucleus by considering 
the ratio of dilepton cross sections in $A-A$ and $p-A$ collisions at RHIC or
LHC. 

\section*{Acknowledgment}
We would like to thank A. Dumitru and R. Fries for useful discussions.
F.G. is supported by CNRS. J. J-M. is supported in part by a PDF from 
BSA and by U.S. Department of Energy under Contract No. DE-AC02-98CH10886.

\bibliographystyle{unsrt}

\end{document}